\documentclass[graybox,openany]{svmult}

\usepackage{type1cm}        % activate if the above 3 fonts are                        % not available on your system
\usepackage{makeidx}         % allows index generation
\usepackage{graphicx}        % standard LaTeX graphics tool
\usepackage{multicol}        % used for the two-column index
\usepackage[bottom]{footmisc}% places footnotes at page bottom
\usepackage{amssymb}	% Extra maths symbols
\usepackage{lineno}
\usepackage{xcolor}       

\usepackage{ulem} 
\usepackage{soul}

\usepackage{newtxtext}       % 
\usepackage{newtxmath}       % selects Times Roman as basic font

\makeindex             % used for the subject index
\begin{document}

\title*{Black holes: accretion processes in X-ray binaries}
%\titlerunning{Accretion processes in BHBs} 
%for an abbreviated version of
\author{Qingcui Bu$^{\star}$ and Shuangnan Zhang}
\institute{Qingcui Bu \at Institut f\"ur Astronomie und Astrophysik, Kepler Center for Astro and Particle Physics, Eberhard Karls Universit\"at, Sand 1, 72076 T\"ubingen, Germany, \email{bu@astro.uni-tuebingen.de}
\and Shuangnan Zhang \at Key Laboratory of Particle Astrophysics, Institute of High Energy Physics \& University of Chinese Academy of Sciences, Chinese Academy of Sciences, 19B Yuquan Road, Beijing 100049, China \\
$\star$ corresponding author
}
\maketitle

%\tableofcontents

\abstract{Accretion onto black holes is one of the most efficient energy source in the Universe. Black hole accretion powers some of the most luminous objects in the universe, including quasars, active galactic nuclei, tidal disruption events, gamma-ray bursts, and black hole X-ray transients. In the present review, we give an astrophysical overview of black hole accretion processes, with a particular focus on black hole X-ray binary systems. In Section \ref{sec1}, we briefly introduce the basic paradigms of black hole accretion. Physics related to accretion onto black holes are introduced in Section \ref{sec2}. Models proposed for black hole accretion are discussed in this section, from the Shakura-Sunyaev thin disk accretion to the advective-dominated accretion flow. Observational signatures that make contact to stellar-mass black hole accretion are introduced in Section \ref{sec3}, including the spectral an d fast variability properties. A short conclusion is given in Section \ref{sec4}.}


\pagenumbering{Roman}
\pagenumbering{arabic}

\setcounter{page}{1}
\thispagestyle{empty}

\section{Introduction}
\label{sec1}

Accretion is the process that describes the inflow of matter toward a central gravitational object or toward the center of an extended system. When the central object is a black hole (BH), the most spectacular observational manifestations of accretion occur. The enormous gravitational field attracts the surrounding matter that tends to be swallowed into the black hole's event horizon (a spherical radius of $\sim GM/c^2$). Since BHs are so small in size compared to the scale of their feeding sources, and the centrifugal forces on the in-falling matter of given angular momentum $j$ increase more rapidly ($\propto R^{-3}$) than gravity ($\propto R^{-2}$) as it moves inward in radius $R$, accretion process is generally believed involving rotational flows. Matter in such flow must lose angular momentum in order to spiral inward the BH and release its gravitational binding energy, leading to bright radiation from the vicinity of the black hole, travelling through the universe, and eventually reaching our telescopes. Therefore, while black holes themselves are the 'black', accretion normally makes the regions directly outside the BH event horizon, some of the most luminous stable compact sources known.
 
Black hole accretion is one of the most ubiquitous processes in astrophysics and is responsible for some of the most luminous objects observed around us, including quasars, active galactic nuclei (AGN), tidal disruption events (TDE), gamma-ray bursts (GRB), and black hole X-ray binaries (XRBs). Black holes come in many varieties, from objects with stellar masses of several times $M_\odot$ to supermassive holes with masses of $10^6 - 10^9 M_\odot$. Particular interestingly, the $1/\nu_{K} \sim \sqrt{M}$ ($\nu_{K}$ is the Keplarian orbital frequency and M is the BH mass) relation scaling adds ``gravity" to observations of stellar mass black holes, because the same processes that might take hundreds of years in quasars, takes only minutes or hours in stellar-mass BHs. Many stellar-mass BHs are found in close X-ray binary systems, in which the primary BH, with a mass from several to tens of $M_\odot$, accretes matter from the secondary companion star. In the present review, we will discuss black hole accretion processes in those black hole X-ray binaries systems and their applications to interpret the observed spectral and timing properties from them.

%%%%%%%%%%%%%%%%%%%%%%%%%%%%%%%%%%%%%%%%%%%%%%%%%%%%%%

\section{Physics of accretion onto BHs} 
\label{sec2}

\subsection{Formation of accretion disk}
\label{sec2:1}

Astrophysical accreting black holes, are simple to describe mathematically, and can be completely described by just two parameters: their mass $M$ and angular momentum $j$. In the simplest case, a non-spinning Schwarzschild black hole, the system can is characterized only by $M$. In a black hole binary system, the companion star is orbiting the BH and when the strong gravitational force of the BH accretes gas that comes from the companion star, the in-falling gas will have some original angular momentum. When the angular momentum of the in-falling gas $j \gg R_{\rm{g}}c$, where $R_{\rm{g}}$ is the gravitational radius of the black hole, an accretion disk forms around the BH. After the gas enters the accretion disk, its gravitational binding energy releases by falling into lower orbits, providing a large amount of radiation that reaches the X bands in the inner regions. 

The radial in-falling of the gas is stopped by the centrifugal potential barrier at a radius $R \sim j^2/R_{\rm{g}}c^2$ (known as the last stable orbit), where the gas cools and forms a ring rotating at Keplerian velocity. Further accretion is possible only when the angular momentum is redistributed, so that some gas could shed enough angular momentum and spiral inward the BH. Because of the conservation of angular momentum, if some gas slows down to spiral inward, some other must transport outward, gradually creating a flatten disk from the initial ring. The disk edge expands with more accretion, far beyond the initial radius, and most of the original angular momentum is carried out to this edge.

In other words, accretion disk is a rotational flow with viscous processes, while the orientation of the accretion disk is defined by the angular momentum of the in-falling gas. Until now, the nature of angular momentum transport mechanism and the processes whereby gravitational binding energy is converted into observable forms of radiation remain the two key questions in black hole accretion theory.

\subsection{Viscous process}
\label{sec2:2}

Several mechanisms have been proposed on the angular momentum extraction for black hole accretion theory. The astrophysical consensus on these mechanisms is that the angular momentum transport in accretion disks largely involves magnetic fields. Theoretically, even a weak magnetic field can destabilize the gas orbital motion and cause the flow to turbulent through the so-called Magneto-rotational instability (MRI) (see \cite{Balbus2004} for review). 

The MRI is a robust prediction of magneto-hydrodynamic (MHD) models, which is defined for plasma that is sufficiently ionized and not too strongly magnetized \cite{Balbus1998}. This instability takes a weak magnetic field inside a accretion disk and amplifies it exponentially until the system becomes non-linear and develops MHD turbulence. Turbulent pulsations force each gas element to diffuse from one Keplerian orbit to another, while the Maxwell and Reynolds stresses in the turbulent state transport angular momentum outward, leading the gas to accrete inward. This process is accompanied by viscous heating that allows gravitational binding energy to be dissipated into heat and radiated away from the disk. 

In order to allow the angular momentum transport stresses outward, a viscosity law must be specified. The viscosity in astrophysical accretion disks can not come from the ordinary molecular viscosity since the value is too weak to provide necessary level of stresses we observed. Until today, the Shakura and Sunyaev '$\alpha$-viscosity' prescription is still the most notable prescription used to explain the viscous stress in accretion disk \cite{Shakura1973}. 

This assumption is based on dimensional arguments, in which the shear stresses $t_{R\Phi}$ are assumed to be directly proportional to the total pressure $P$. The kinematic viscosity coefficient $\nu$ is then parameterized from,
%----------------------------------------------
\begin{equation}
\label{eq.:alpha}
  \nu \sim l_0\upsilon_0,
\end{equation}
%-------------------------------------------------------------
where $l_0$ is the correlation length of turbulence and $\upsilon_0$ is the mean speed of turbulence. In realistic situations, $\upsilon_0$ can not exceed the sound speed $c_s$ and $l_0$ can not be larger than the height of disk $H$, Eq. \ref{eq.:alpha} then becomes, 
%----------------------------------------------
\begin{equation}
\label{eq.:alpha2}
  \nu \equiv \alpha c_s H = \alpha\frac{c^2_s}{\Omega_K},
\end{equation}
%-------------------------------------------------------------
where the dimensional coefficient $\alpha$ is assumed to be a constant, varying between 0 and 1. The shear stresses thereby has a form of $t_{R\Phi} = \alpha P$. $\alpha$ has a typical value of $\sim 0.02$ according to the estimation from magneto-hydrodynamic (MHD)simulations \cite{Hawley2011}, however observations suggest a value of $\sim 0.1$ \cite{King2007}.

The $\alpha$ prescription allows models to be built that couple the dynamics of the flow to the thermodynamics of it, phenomenologically, without addressing the fundamental physics. When combined with extra radiative cooling processes, one can get a full set of models that uses to explain the energy spectra, time variability, etc. we observe. 

The non-linear behavior of the MRI and the turbulence it generates are complicated and can only be studied by numerical simulations. Plenty of simulations have been done in the limit of a local shearing box that supports high spatial resolution (see \cite{Balbus2004} for review). These studies suggest that the MHD is inevitable as long as the gas and the magnetic field are well-coupled and that the Maxwell stress dominates over the Reynolds stress by several factors. 

The MRI turbulence mechanism is widely applied to the sources whose power output is dominated by thermal radiative emissions. The dynamics of the accretion disk then depends critically on how much this thermal energy is radiated away (Section \ref{sec2:5}). By far, the MRI turbulence is probably the only mechanism that can be described by the classical '$\alpha$-viscosity' prescription. However, so far, most of the research has focused on the dynamic of the MRI turbulence, not the thermodynamics that directly links to the observations, which leaves us a lot of unanswered fundamental questions of thermodynamics. 

%1) \textit{External stresses associated with large scale magnetic fields in a magneto-hydrodynamic (MHD) outflow} \cite{Blandford1982}. This mechanism is usually related to the low luminosity sources, where accretion power might be largely converted into mechanical power in outflows. (2) \textit{Magneto-rotational instability (MRI) turbulence}. This process is widely applied to the sources whose power output is dominated by thermal radiative emissions. The dynamics of the accretion disk then depends critically on how much this thermal energy is radiated away (Section \ref{sec3:2}). (3)\textit{\mod{Non-axisymmetric} waves and shocks}. Such waves can transport angular momentum outward through the accretion flow. This process could play an important role in the outer regions of disk due to the tidal excitation by the companion star or in the inner region of accretion flows whose angular momenta are misaligned with the black hole spin axis \cite{Fragile2008, Balbus1999}.

\subsection{Fundamental Principles}

Before getting into details, let us briefly overview the basic parameters available in BHs. There are three generic types of physical processes that must happen in BH accretion disks: 'dynamical' processes with a characteristic timescale, $t_{\rm{dyn}} \sim 1/\Omega$ , where $\Omega$ is the orbital angular velocity; 'thermal' processes with a characteristic time-scale, $t_{\rm{th}} \sim c^2_{s}/\nu \Omega^2$, where $c_s$ is the sound speed and $\nu$ is the kinematic viscosity coefficient; and 'viscous' processes with a characteristic time-scale, $t_{\rm{vis}} \sim R^2/\nu$ , where $R$ is the radial distance from the black hole. Typically, dynamical processes move much faster than thermal or viscous ones,
%------------------------------------------------------------
\begin{equation}
\label{eq.timescales}
  t_{\rm{dyn}} \ll t_{\rm{th}} \ll t_{\rm{vis}} .
\end{equation}
%------------------------------------------------------------

Generally, in the first approximation, one may only consider the dynamical structure. The dynamical structure of the disk is sculptured by three forces: gravity, pressure and rotation forces. The relative importance of each of the three forces leads to different dynamical states of accretion (thin, slim, thick, ADAF), which will be introduced in the Section \ref{sec2:4}.

To obtain a steady-state disk, the dynamical behavior of the gas must follow the fundamental conservation laws of the rest mass, energy and angular momentum \cite{Abramowicz1988, Narayan1998}, 
%------------------------------------------------------------
\begin{equation}
\label{eq.:conservation1}
  \frac{\mathrm{d}}{\mathrm{d} R} (\rho RH \upsilon) = 0,
\end{equation}

\begin{equation}
\label{eq.:conservation2}
  \upsilon\frac{\mathrm{d} \nu}{\mathrm{d} R} - {\Omega}^2R = -\Omega^2_{K}R - \frac{1}{\rho}\frac{\mathrm{d}}{\mathrm{d} R}(\rho c^2_s),
\end{equation}

\begin{equation}
\label{eq.:conservation3}
  \upsilon\frac{\mathrm{d} (\Omega R^2)}{\mathrm{d} R} = \frac{1}{\rho RH}\frac{\mathrm{d}}{\mathrm{d} R}(\nu \rho R^3H \frac{\mathrm{d} \Omega}{\mathrm{d} R}),
\end{equation}

\begin{equation}
\label{eq.:conservation4}
  \rho \upsilon(\frac{\mathrm{d} e}{\mathrm{d} R} - \frac{P}{\rho^2}\frac{\mathrm{d} \rho}{\mathrm{d} R}) = \rho\nu R^2(\frac{d\Omega}{dR})^2 - q^{-},
\end{equation}
%-------------------------------------------------------------
where $\rho$ is the density of the gas, $R$ is the radius, $H$ is the vertical scale height, $\Omega_K$ is the Keplerian angular velocity, $\upsilon$ is the radial velocity, $c_s \equiv P/\rho$ is the sound speed, $P$ is the pressure, $e$ is the specific internal energy, $\nu$ is the kinematic viscosity coefficient, and $q^{-}$ is the radiative cooling rate per unit volume.

Equations \ref{eq.:conservation1}--\ref{eq.:conservation4} can generally describe all accretion structures in steady-states, including the Shakura-Sunyaev disk, the slim disk and the Advective-dominated accretion flows (Section \ref{sec2:4}). These conservation equations should be modified when there is an outflow \cite{Poutanen2007, Xie2008} or the mass accretion rate $\dot{m}$ is not constant. 

\subsection{Accretion disk models}
\label{sec2:4}

Besides the angular momentum transport mechanism, the other key question in black hole accretion theory concerns how the gravitational energy converts to the observable radiation, known as the radiative processes and radiative transfer. Generally, these depend on the thermodynamic state of the gas, involving its electron density, ion density, temperature, motion, magnetic filed, and most importantly the optical depth $\tau$ that decides whether the accretion disk is efficiently cooled by radiation. Therefore, in steady states, the structure of the accretion flow is derived by balancing viscous heating against radiative cooling.

The radiative efficiency is defined as $\epsilon \equiv L/\dot{m}c^2$, where $L$ is the disk luminosity, and $c$ is the speed of light. The stead-state accretion flow can be completely described by the mass $M$ and angular momentum $j$, together with the mass accretion rate $\dot{m}$. Subsequently, the key parameters in the process of accretion onto a black hole are the angular momentum of the in-falling gas and the radiative efficiency $\epsilon$, or accretion rate $\dot{m}$. 

The accretion flow around BHs can thereby be classified into four major classes on the basis of their radiative efficiency $\epsilon$ and accretion rate $\dot{m}$. In this section, we will briefly discuss the basic forms of accretion flow in accreting black holes.

\subsubsection{Shakura-Sunyaev disks}
\label{sec2:4:1}

If the cooling time of the gas is shorter than the gas takes to flow into the BH, the gas cools rapidly and forms a geometrically thin accretion disk, also known as the Shakura-Sunyaev disk (SSD). The radaitive efficiency $\epsilon$ is $\sim 0.06$ to $\sim$0.1, meaning that all the heat generated by viscosity at a given radius is immediately radiated away from the disk surface. This regime is well recognized by the thermal emission from the soft state of black hole X ray binaries (see reviews by \cite{Falanga2015}, as well as from some luminous AGNs, such as narrow line Seyfert galaxies and numerous radio quiet quasars (\cite{Foschini2015, Blandford2019}).

The SSD model applies when the $\dot{m}$ is below the Eddington accretion rate $\dot{m}_{\rm{Edd}} \equiv L_{\rm{Edd}}/(\epsilon c^2)$, where $L_{\rm{Edd}} \equiv 4\pi GMc/\kappa$. $M$ is the mass of the black hole and $\kappa$ is the electron scattering opacity, usually taken to be 0.4 $\rm{cm^2/g}$. In such models, the viscous torques transport its angular momentum to the outer part of the disk, while the associated viscous stresses generate heat that is radiated locally from the disk surfaces. The gas cools efficiently and the sound speed $c_s$ is much smaller than the local Keplerian speed $\nu_K$. 

The SSD solution assumes that all physical quantities can be decided by the two spatial coordinates: the radial distance $R$ from the BH and the vertical distance $H$ from the equatorial plane. The dynamical behavior of the gas follows the conservation laws given by Eq. \ref{eq.:conservation1} -- \ref{eq.:conservation4}. The Shakura-Sunyaev disk assumes that the mass accretion rate $\dot{m}$ is constant with radius $R$,
%-------------------------------------------------------------
\begin{equation}
    \dot{m} = 2\pi R \sum(R) = constant,
\end{equation}
%--------------------------------------------------------------------
where $\sum(R)$ is the surface mass density of the disk. The $\dot{m}$ is a free parameter of the model. In the following, we will explore the thin disk structures in Shakura-Sunyaev disk (SSD) solution.

In the geometrically thin disk limit ($H \ll R$), the hydrodynamic equations decouple in the radial and vertical directions. In the radial direction, the Keplarian velocity introduces shear stresses that dissipate heat, which produces a local radiative flux that leaves the disk surface vertically from the top to bottom. We have:
%-------------------------------------------------------------
\begin{equation}
\label{eq.F_rad}
    F_{\rm{rad}} = \frac{3}{8\pi}\frac{GM\dot{m}}{R^3} \left [1 - (R_{\rm{ISCO}}/R)^{1/2} \right ],
\end{equation}
%--------------------------------------------------------------------
assuming that the viscous torque vanishes at the innermost stable circular orbit (ISCO), from where the gas falls free into the BH and no radiation comes out. Here $R_{\rm{ISCO}} = 6R_{\rm{g}}$ for a Schwarzschild BH, and $R_{\rm{g}} \le R_{\rm{ISCO}} \le 9R_{\rm{g}}$ for a Kerr BH, where $R_{\rm{g}} = GM/c^2$ is the gravitational radius. Then the total luminosity emitted by both surfaces of the disks is 
%-------------------------------------------------------------
\begin{equation}
\label{eq.L_disk}
   L =  \int_{R_{\rm{ISCO}}}^{\infty} 2F_{\rm{rad}} \times 2\pi R \mathrm{d} R = \frac{GM\dot{m}}{2R_{\rm{ISCO}}} = \frac{R_{\rm{g}}}{2R_{\rm{ISCO}}} \dot{m}c^2. 
\end{equation}
%--------------------------------------------------------------------
The radiative efficiency $\epsilon \equiv L/\dot{m}c^2 = R_{\rm{g}}/2R_{\rm{ISCO}}$ is $8.3\%$ for a Schwarzschild BH and $50\%$ for an extreme-spin Kerr BH. Note that these values are a bit smaller ($\sim 4\%$) when considering general relativistic corrections \cite{Shapiro1983}. At smaller $R$, more luminosity is generated, which is dissipated over a smaller area, leading an increase in the disk temperature. 

In the optically thick limit ($\tau \gg 1$), in the vertical direction, the gas and radiation are in local thermodynamic equilibrium, the radiative flux from the disk surfaces (Eq. \ref{eq.F_rad}) can be written in terms of the $F_{\rm{rad}} = \sigma T_{\rm{d}}^4$ \cite{Davis2006, Dunn2011}, where $\sigma$ is the Stephan-Boltzmann constant and $T_{\rm{d}}$ is the surface temperature of the disk. Thus, one gets
%------------------------------------------------------
\begin{equation}
\label{eq.:T_disk}
    T_{\rm{d}}(R) = (\frac{F_{\rm{rad}}}{\sigma})^{1/4} = 10^5 M_{8}^{1/4} (\frac{\dot{m}}{\dot{m_{\rm{Edd}}}})^{1/4} (\frac{R}{R_{\rm{g}}})^{-3/4} \left [1 - (R_{\rm{ISCO}}/R)^{1/2} \right ]^{1/4} K,
\end{equation}
%--------------------------------------------------------------- 
and this is valid regardless of the disk physics as long as the disk is optically thick. When neglecting the inner boundary conditions allowing easy integration of the total flux (as in the \textsc{diskbb} model \cite{Mitsuda1984}), the flux peaks at $E_{\rm{peak}} \sim 2.36 \kappa T_{\rm{in}}$, where $\kappa$ is the electron scattering opacity and $T_{\rm{in}}$ is the temperature at the inner edge of the disk. %$E_{\rm{peak}}$ for stellar-mass BH is $\sim$ 1--10 keV in the X-ray band.

The SSD described above is a global solution, which assumes that the mass accretion rate $\dot{m}$ is constant with radius $R$ and the viscosity is proportional to the total pressure or only the gas pressure ('$\alpha$-viscosity', Section \ref{sec2:2}). However, a constant $\dot{m}$ is not always true, the SSD is challenged by two major instabilities, one relates to the ionization of hydrogen which is triggered at relatively low luminosity, and the other relates to the radiation pressure that likely occurs at higher luminosity (see \cite{Done2007} for review). 

%This regime is well recognized by the thermal emission from the soft state of black hole XRBs (see reviews by \cite{Abramowicz2013, McClintock2013, Falanga2015}, as well as from some luminous AGNs, such as narrow line Seyfert galaxies and numerous radio quiet quasars (\cite{Foschini2015, Blandford2019}).

\subsubsection{Advective dominated accretion flows}
\label{sec2:4:2} 

On the contrary to the cold disk is the hot accretion flow solution. When the $\dot{m}$ is much lower than in the SSD, with $L \ll 0.1 \dot{m}c^2$, the system switches to a different accretion regime, known as the Advection Dominated Accretion Flow (ADAF). In such models, the gas cooling time is much longer than the accretion time and nearly all of the viscous dissipated energy is advected into the BH rather than radiated away \cite{Narayan1994, Narayan1995}. Because of the inclusion of the advection, the accretion gas is thermally stable, which is the key characteristic of the ADAF solutions. This regime is widely applied to the emission observed from the low luminosity XRBs and AGNs (see \cite{Yuan2014} for review).

In ADAF solutions, the gas pressure is large, which consequently causes the accretion gas become geometrically thick, optically thin, and quasi-spherical. Besides, the considerable pressure-support in the radial direction makes the angular velocity $\Omega$ become sub-Keplarian. The radial velocity $\upsilon$ is relatively large $\sim \alpha\upsilon_{k}(H/R)^2 \sim \alpha c/R^{1/2}$, which in turn contributes to a short accretion time $t_{\rm{acc}} = R/\upsilon \sim R^{3/2}/\alpha c$. Because of the small optical depth $\tau$, the emitted radiation is almost never blackbody, but instead is dominated by synchrotron, bremsstrahlung, and inverse Compton scattering (Section \ref{sec2:5}).

Like all matter, the hydrodynamics of the accretion gas follows the conservation laws given by Eq. \ref{eq.:conservation1} -- \ref{eq.:conservation4}. Note, unlike a constant $\dot{m}$ in SSD, the $\dot{m}$ in ADAFs decreases with decreasing $R$, thus the conservation equations Eq. \ref{eq.:conservation1} -- \ref{eq.:conservation4} should be modified \cite{Poutanen2007, Xie2008, Poutanen2013}. 

Assuming a power-law variation between $\dot{m}$ and $R$ for simplicity, Eq. \ref{eq.:conservation1} becomes \cite{Blandford1999}
%------------------------------
\begin{equation}
\label{eq.ADAF_M_dot}
    \dot{m}(R) = 4 \pi \rho RH |\upsilon| = \dot{m}_{\rm{BH}}(\frac{R}{R_S})^{s},  (R_S \le R \le R_{\rm{out}}; 0 \le s \le 1),
\end{equation}
%-------------------------------------
where $R_S = 2GM/c^2$ is the Schwarzschild radius of the BH, $R_{\rm{out}}$ is the outer radius of the accretion flow. The index $s$ is a measure of the strength of the outflow, whereas $s=0$ equals to a constant $\dot{m}$ without outflow. Assuming that the only important change the mass outflow brings is in the density profile \cite{Xie2008}, the self-similar solution of Eq. \ref{eq.:conservation2} -- \ref{eq.:conservation4} and \ref{eq.ADAF_M_dot} becomes approximately \cite{Yuan2014} 
%------------------------------------------
\begin{equation}
\label{eq.ADAF_V}
    \upsilon \approx -1.1 \times 10^{10} \alpha r^{-1/2} \rm{cm}/\rm{s},
\end{equation}

\begin{equation}
\label{eq.ADAF_omega}
    \Omega \approx 2.9 \times 10^4 m^{-1} r^{-3/2} \rm{s}^{-1},
\end{equation}

\begin{equation}
\label{eq.ADAF_sp}
    c_{s}^{2} \approx 1.4 \times 10^{20} r^{-1} cm^2 \rm{s}^{-2},
\end{equation}

\begin{equation}
\label{eq.ADAF_ne}
    n_e \approx 6.3 \times 10^{19} \alpha^{-1}m^{-1} \dot{m}_{\rm{BH}} r^{-3/2+s} \rm{cm}^{-3},
\end{equation}

\begin{equation}
\label{eq.ADAF_B}
    n_e \approx 6.5 \times 10^8 (1+\beta)^{-1/2} \alpha^{-1/2}m^{-1/2} \dot{m}_{\rm{BH}}^{-1/2} r^{-5/4+s/2} \rm{G},
\end{equation}

\begin{equation}
\label{eq.ADAF_p}
   P \approx 1.7 \times 10^{16} \alpha^{-1}m^{-1} \dot{m}_{\rm{BH}} r^{-5/2+s}  \rm{g} \rm{cm}^{-1} \rm{s}^{-2},
\end{equation}
%-------------------------------------------
where the BH mass $M$, the mass accretion rate $\dot{m}$, and the radiu $R$ are scaled to solar, Eddington, and Schwarzschild unites, respectively:
%-------------------------------------------
\begin{equation}
    m \equiv M/M_{\odot}, \dot{m} \equiv \dot{m}/\dot{m_{\rm{Edd}}}, r \equiv R/R_{\rm{S}}.
\end{equation}
%------------------------------------------
The $\beta \equiv P_{\rm{gas}}/P_{\rm{mag}}$ is a measure of the magnetic filed strength \cite{Narayan1995}, where $P_{\rm{mag}} \equiv B^2/8\pi$ is the magnetic pressure and $P_{\rm{gas}}$ is the gas pressure.

When it comes to the thermodynamics of ADFAs, one generally assumes that the accretion gas has a two-temperature plasma (at least at small radii), in which the electron temperature $T_e$ is much less than iron temperature $T_i$. In order for the gas to be two-temperature, there must be a coupling between electrons and irons, which is generally assumed to be occurred via Coulomb collisions. For such plasmas, the energy equation (Eq. \ref{eq.:conservation4}) is modified into two coupled equations (\cite{Nakamura1997, Quataert1999a, Yuan2014},
%--------------------------------
\begin{equation}
\label{eq.:adv,i}
    q^{\rm{adv},i} \equiv \rho \upsilon (\frac{\mathrm{d}e_i}{\mathrm{d} R} - \frac{p_{i}}{\rho^2}\frac{\mathrm{d}\rho}{\mathrm{d}R}) \equiv \rho \upsilon \frac{\mathrm{d}e_i}{\mathrm{d} R} - q^{i,c} = (1- \delta) q^{+} - q^{ie},
\end{equation}
 
\begin{equation}
\label{eq.:adv,e}
    q^{\rm{adv},e} \equiv \rho \upsilon (\frac{\mathrm{d}e_e}{\mathrm{d} R} - \frac{p_e}{\rho^2}\frac{\mathrm{d}\rho}{\mathrm{d}R}) \equiv \rho \upsilon \frac{\mathrm{d}e_e}{\mathrm{d} R} - q^{e,c} = \delta q^{+} + q^{ie} - q^{-}, 
\end{equation}
%-------------------------------
where $e_i$ and $e_e$ are the internal energies of irons and electrons per unit mass of the gas, $p_i$ and $p_e$ are the respective pressure. $q^{+}$, $q^{-}$ and $q^{\rm{adv}}$ are the heating rate, cooling rate and advective cooling rate per unit volume, respectively, with a assumption of $q^{\rm{adv}} = q^{+} - q^{-}$. The quantity $q^{ie}$ is the rate of transfer of thermal energy from ions to electrons via Coulomb collisions. $\delta$ is the fraction of electron heating to total heating, while the rest (1 - $\delta$) goes into the ions heating. Usually, $\delta$ is treated as a free parameter in ADAFs solution. Note, Eq. \ref{eq.:adv,i} -- \ref{eq.:adv,e} need further modified if one includes the contribution of magnetic fields \cite{Quataert1999b}.

Plenty of work has been done to constrain the value of $\delta$ from first principles by considering the effects of magnetic re-connection  MHD turbulence, or dissipation of pressure anisotropy in a collision-less plasma (see \cite{Werner2017} for review). Despite that no consensus has been reached, most work generally suggest that $\delta \gg 0.01$. 

By modelling astrophysical observations of hot accretion flows in AGNs and XRBs \cite{Yuan2003, Yu2010, Liu2013, Yang2015, Xie2016}, researchers have obtained new constraints on the value of $\delta$. These work suggest that $\delta$ probably lie in the range 0.1--0.5. It should be noted that, even if electrons and ions receive equal amounts of the dissipated energy ($\delta = 0.5$), the plasma can still be two-temperature \cite{Narayan2008}.

At this point, a global solution of ADAF can be obtained by solving the modified equations, e.g., Eq. \ref{eq.ADAF_M_dot}, \ref{eq.:conservation2}, \ref{eq.:conservation3}, \ref{eq.:adv,i} and \ref{eq.:adv,e}. Mathematically, the global solution must follow three boundary conditions: (1) there has to be an intermediate sonic radius $R_{\rm{sonic}}$ where $\upsilon = c_s$, since the radial velocity of the accreting gas is highly subsonic at large radius, (2) the viscous torque must disappear at the ISCO as in SSD, (3) and the properties of the accretion flow should be consistent with the in-falling gas from the outside. Therefore, the main parameters remain here are BH mass $M$, mass accretion rate $\dot{m}_{\rm{BH}}$, magnetization parameter $\beta$, wind parameter $s$, viscosity parameter $\alpha$, and electron heating parameter $\delta$. Among these parameters, $\delta$ is the only free parameter, the other parameters can either estimate through observations ($M$, $\dot{m_{\rm{BH}}}$) or obtain from numerical simulations ($\alpha$, $\beta$, $s$ ). 

\begin{figure}
\includegraphics[width=10cm]{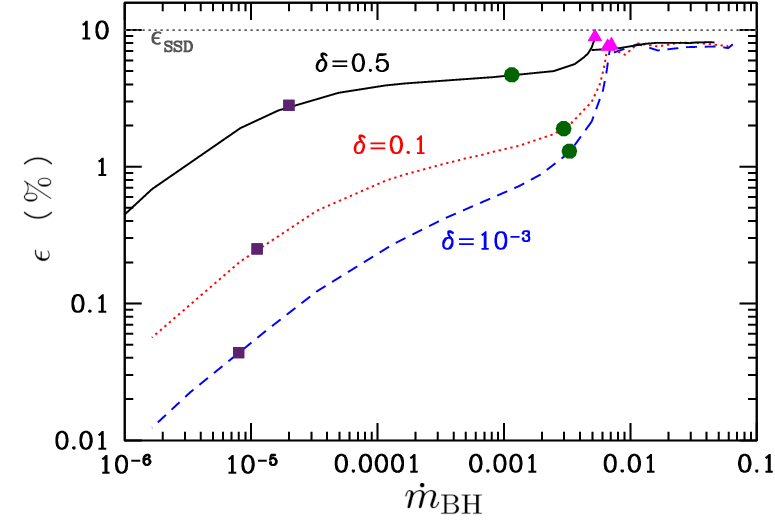}
\caption{Radiative efficiency $\epsilon$ of a luminous hot accretion flow as a function of the BH mass accretion rate $\dot{m_{\rm{BH}}}$ for three values of the electron heating parameter $\delta$, with a set of parameters fixed as $\alpha = 0.1, \beta = 9, s = 0.4$. The nominal radiative efficiency of a standard thin disk, $\epsilon_{\rm{SSD}} = 10\%$, is indicated by the horizontal dashed line at the top. Figure reproduced from Yuan \& Narayan (2014)}
\label{fig1} 
\end{figure}

Fig. \ref{fig1} shows the radiative efficiency $\epsilon$ as a function of the mass accretion rate $\dot{m_{\rm{BH}}}$ for various values of the electron heating parameter $\delta$. For a given $\delta$, the radiative efficiency of an ADAF increases with increasing mass accretion rate and approaches the efficiency of the SSD ($\epsilon_{\rm{SSD}} \sim 0.1$) at high accretion rate. Besides, the radiative efficiency strongly depends on the assumed value of $\delta$.
%-----------------------------------------------------------------
%--------------------------------------------------------
Once we know the values of all parameters, the global solution can give the radial distributions of the parameters in Eq. \ref{eq.ADAF_V} -- \ref{eq.ADAF_p}, together with the intermediate sonic radius $R_{\rm{sonic}}$. Fully relativistic solutions of ADAFs corresponding to the Kerr metric have also been found numerically \cite{Abramowicz1996, Peitz1997, Gammie1998, Manmoto2000}. Further discussion of ADAFs can be find in the review articles from Yuan and Narayan \cite{Narayan2008, Yuan2014}.

\subsubsection{Slim disks}
\label{sec2:4:3}

When the $\dot{m}$ approaches or exceeds $\dot{m}_{\rm{Edd}}$, the accreted gas on the disk becomes optically too thick to radiate all the energy locally, meaning that the viscous heating can not be balanced by the radiative cooling locally thus triggers another cooling mechanism, i.e., advection. The radiative efficiency $\epsilon$ is less than $\sim$0.1. Consequently, a thin disk becomes a slim disk \cite{Abramowicz1988} (also, see \cite{Czerny2019} for a review). The slim disk model is widely applied to some ultra-luminous X-ray sources and galactic X-ray binaries in super-Eddington states \cite{Watarai2001, Sadowski2009}. 

Both the slim disk and thin SSD disk have optically thick, cold accretion flows, and are based on the '$\alpha$-viscosity' assumption. In some sense, the slim disks are more physical than SSDs, since they are assumed to extend down to the BH horizon. The local emission is no longer given by the local dissipation, and a fraction of energy is lost through the advection of the heat. The radiative efficiency $\epsilon$ is much lower than in the SSD and decreases with the increasing mass accretion rate $\dot{m}$. 

The slim disk solutions were first introduced in the pseudo-Newtonian limit by Abramowicz \cite{Abramowicz1988}, and can be modified from Eq. \ref{eq.:conservation1} -- \ref{eq.:conservation4} with reasonable advection term $q_{\rm{adv}}$, radial pressure gradients, and proper boundary conditions. Note, different from the SSD/ADAF, the viscous torque should vanish at the BH horizon instead of the ISCO. 

\begin{figure}
\includegraphics[width=55mm]{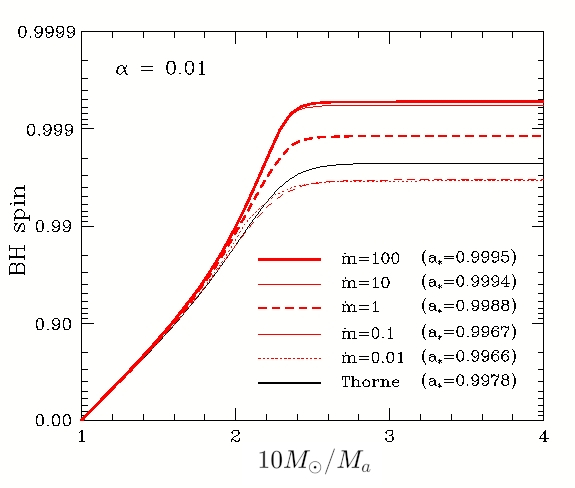} 
\includegraphics[width=55mm]{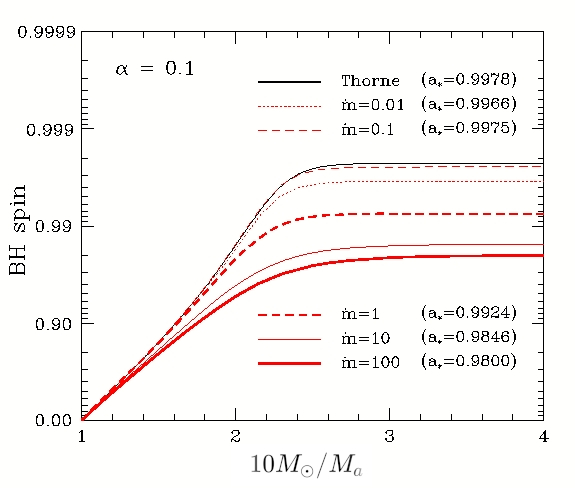}
\caption{Profiles of the BH spin for $\alpha$ = 0.01 (left) and $\alpha$ = 0.1 (right panel) for different accretion rates $\dot{m}$. $M_{a}$ is the accreted mass. The red lines show the results for different accretion rates, while the black line indicates the classical Thorne solution based on the Novikov \& Thorne (1973) model of thin accretion disk \cite{Novikov1973}. Figure adapted from Sadowski (2011) \cite{Sadowski2011}.}
\label{fig2} 
\end{figure}

Further numerical development of slim disk solutions in full General Relativity gives the opportunity to study the main parameters on the BH spin in X-ray binaries \cite{Sadowski2009, Sadowski2011} (see Fig. \ref{fig2}). Their results suggest that the BH equilibrium spin value depends strongly on the assumed value of viscosity coefficiency $\alpha$. In particular, when $\alpha = 0.01$, the BH spin increases with an increasing $\dot{m}$. However, at $\alpha = 0.1$, the BH spin shows an opposite trend with an increasing $\dot{m}$. Besides, the low $\dot{m}$ limit ($\dot{m} = 0.01$) does not perfectly agree with the the classical Thorne solution (black line). This could be because that the slim disk model does not account for the angular momentum carried away by radiation.
%----------------------------------------------------
%------------------------------------------------

One of the general properties of slim disk is that it may rotate with an angular momentum profile significantly different than the Keplerian one, the higher the accretion rate $\dot{m} = \dot{m}c^2/16L_{\rm{Edd}}$, the more significant the departure. 

Despite high super-Eddington accretion rates, the slim disk luminosity remains moderately super-Eddington at a value of about 10 times higher than the Eddington luminosity \cite{Sadowski2009}. This could be because that the geometry of the flow is not spherical and most of the accretion occurs in the equatorial plane while the radiation escapes vertically, hence the radiation can not stop the inflow (see \cite{Abramowicz2013} for a review).

When the $\dot{m} \gg \dot{m}_{\rm{Edd}}$, there is so much matter in the accretion flow that the radiation is trapped and can not escape before falling into BH. Thus the in-falling gas has a very low radiate efficiently of $\epsilon \ll 0.1$. The accretion disk becomes geometrically and optically thick, known as the Polish doughnuts \cite{Paczynsky1980, Abramowicz1989, Narayan1998}. Usually, Such high $\dot{m}$ can never be reached for accreting black holes in XRBs or AGNs, where characteristic rates are below the $\dot{m}_{\rm{Edd}}$. Therefore, in Section \ref{sec3}, we will focus on the interpretations of the thin disk, slim disk and ADAFs in XRBs. 

\subsubsection{Disk-corona and jets}
\label{sec2:4:4}

Corona is essentially an energetic plasma (hot electrons), which is generally believed to be located in the vicinity of the BH. There is no commonly accepted mechanism of the formation of corona. 

One of most accepted mechanism is the evaporation of the disk under the heat conduction \cite{Meyer1994}. In this model, corona forms above the thin disk either by processes similar to those operating in the surface of the sun, or by a thermal instability in the uppermost layers of the disk. It has been a success recent years in explaining both the physical picture of the corona formation and the truncation of the disk and the observed X-ray spectra of XRBs (see \cite{Liu2022} for a review).

Another plausible mechanism involves magnetic filed. The idea is that a low density corona could be heated by reconnecting magnetic loops emerging from the accretion disk \cite{Galeev1979}. Due to a combination of the differential Keplerian rotation and the turbulent convective motions in the disk, a seed magnetic field can be exponentially amplified with a timescale of $t_{\rm{G}} \sim R/3\nu_{c}$ at a radius of $R$, where $\nu_{c}$is the convective velocity. As previously introduced in Section \ref{sec2:2}, the MHD turbulence developed by the MRI can efficiently generate the magnetic energy in the disk. The instability operates on a Keplerian time-scale, thus a low dissipation rate inside the disk would lead to buoyant transport of generated magnetic loops to the corona. This 'corona-dominated' dissipation could be used to explain the ADAF state of the accretion disk. By contrast, in SSDs, the bulk of the energy is released inside the optically thick disk and the coronal activity is significantly suppressed.

At last, we will very briefly introduce the jets since there has long been a strong observational connection between accreting black holes and relativistic jets across all scales of black hole mass. For stellar-mass black holes, it is now widely convinced that the jets come from an electromagnetic form of the Penrose process suggested by Blandford and Znajek (BZ) \cite{Blandford1977}. Their model suggested that the jet energy is extracted from the spin energy of the BH via a torque provided by magnetic field lines that thread the event horizon or ergosphere. The model is further supported by the GRMHD simulations of accreting black holes, from which they show MHD jets can form spontaneously from generic initial conditions \cite{McKinney2009}. Especially, when the accretion discs are thick, then stronger jets and winds are driven by either stronger turbulent magnetic fields or by large-scale fields advected from large radii. Numerically, the jet power depends on a black hole's mass $M$ and spin, the mass accretion rate $\dot{m}$, the strength and topology of the magnetic field. Other variables need to be eliminated in order to study the connection between the BH spin and relativistic jet. 

%Two kinds of jets have been identified from the radio emission of the XRBs, which are the 'steady jets' and 'ballistic jets', each relates to a specific spectral state of the source.

\subsection{Radiation Cooling}
\label{sec2:5}

In the Shakura-Sunyaev disk (Section \ref{sec2:4:1}), the radiation efficiency is high and nearly all the heat generated within the disk is radiated locally. The disk remains relatively cold and radiate like a blackbody or modified blackbody, with a diffusion approximation of total optical depth $\tau$ coming from the absorption depth and electron scattering optical depth $\tau = \tau_{abs} + \tau_{es} \gg 1$. The radiation emissivity $f$ (emission rate per unit volume) has a form of 
%-------------------------------------------------
\begin{equation}
\label{eq.f_disk}
    f = \frac{8\sigma{T_e}^4}{3H\tau},
\end{equation}
%----------------------------------------------------
where $\sigma$ is Stefan-Boltzmann constant and $T_e$ is the electron temperature at the equatorial plane. 

In the optically thin disk limit ($\tau \ll 1$), such as ADAFs (Section \ref{sec3:2}), radiation is inefficient, and the accretion disks remain relatively hot and geometrically thick. The relevant radiation processes are synchrotron cooling ($f_{\rm{synch}}$) and bremsstrahlung cooling ($f_{\rm{br}}$), modified by Compton cooling ($f_{\rm{synch,C}}, f_{\rm{br,C}}$).

Thermal bremsstrahlung is caused by the inelastic scattering of relativistic thermal electrons off (non-relativistic) ions and other electrons. The radiation emissivity of bremsstrahlung $f_{\rm{br}}$ includes two parts: the ion-electron part $f_{\rm{ei}}$ and the electron-electron part $f_{\rm{ee}}$. $f_{\rm{br}}$ becomes \cite{Narayan1995}
%-------------------------------------------
\begin{equation}
\small
\label{eq.f_br}
f_{\rm{br}} =
\begin{cases}
n_e \overline{n} \sigma_{T}c\alpha_{f}m_{e}c^2 \times \left [4(\frac{2\theta_{e}}{{\pi}^3})^{1/2}(1 + 1.781\theta_{e}^{1.34})\right ] + \\
n_{e}^{2}cr_{e}^{2}m_{e}c^2\alpha_{f} \times \left [\frac{20}{9\pi^{1/2}}(44 - 3\phi^{2})\theta_{e}^{3/2}(1 + 1.1 \theta_{e} + \theta_{e}^{2} - 1.25 \theta_{e}^{5/2})\right ]&  \theta_{e} < 1, \\

n_e \overline{n} \sigma_{T}c\alpha_{f}m_{e}c^2 \times \left \{ \frac{9\theta_{e}}{2\pi}\left [\rm{ln}(1.123\theta_{e} + 0.48) +1.5\right ]\right \} + \\
n_{e}^{2}cr_{e}^{2}m_{e}c^2\alpha_{f} \times \left \{24\theta_{e}\left [\rm{ln}(1.123\theta_{e}) +1.28\right ]\right \}& \theta_{e} \ge 1,

\end{cases}
\end{equation}
%-------------------------------------------
where $n_{e}$ is the electron number density, $\overline{n}$ is the ion number density averaged over all species, $\sigma_{T}$ is the Thomson cross section, $\alpha_{f} = 1/137$ is the fine structure constant, $\theta_{e} = \frac{\kappa_{B}T_{e}}{m_{e}c^2}$ is the dimensionless electron temperature, $\kappa_{B}$ is the Boltzmann constant and $r_{e} = e^2/m_{e}c^2$ is the classical radius of the electron.

If the accretion environment is threaded by magnetic fields, the hot electrons can also be cooled by synchrotron emission. The relativistic Maxwell distribution of electrons is \cite{Narayan1995}
%---------------------------------------------------------------------------
\begin{equation}
\label{eq.f_syn}
    f_{\rm{synch}} = \frac{2\pi}{3c^2}kT_{e}\frac{\mathrm{d}}{\mathrm{d} R} \left [ \frac{3eB\theta_{e}^{2}x_{M}}{4\pi m_{e}c}\right ],
\end{equation}
%------------------------------------------------------------------------
where $e$ is the electric charge, $B$ is the equipartition magnetic field strength. $x_M$ is the solution of the transcendental equation
%---------------------------------------------------------------------------
\begin{equation}
    \rm{exp}(1.8899x_{M}^{1/3}) = 2.49 \times 10^{-10} (\frac{4\pi n_{e} R}{B})\frac{x_{M}^{-7/6} + 0.4x_{M}^{-17/12} + 0.5316x_{M}^{-5/3}}{\theta_{e}^{3}K_{2}(1/\theta_{e})},
\end{equation}
%------------------------------------------------------------------------
where $R$ must be in physical units and $K_2$ is the modified Bessel function of the second kind. This expression is valid only for $\theta_{e} > 1$, which is sufficient in most cases.

In addition, the hot electrons can also Compton up-scatter the soft photons from bremsstrahlung and synchrotron radiation. The formulae for these are \cite{Narayan1995, Abramowicz2013}
%----------------------------------------
\begin{equation}
\label{eq.f_br,C}
    f_{\rm{br,C}} = f_{\rm{br}}\left \{ \eta_{1} - \frac{\eta_1 x_c}{3\theta_e} - \frac{3\eta_1 \left [ 3^{-(\eta_{3} +1)} - (3\theta_e)^{-(\theta_{3}+1)} \right ]}{\eta_3 + 1} \right \}
\end{equation}

\begin{equation}
\label{eq.f_syn,C}
    f_{\rm{synch,C}} = f_{\rm{synch}}\left [\eta_{1} - \eta_{2}(x_{c}/\theta_{e})^{\eta_3} \right ].
\end{equation}
%-----------------------------------------------------------
Here $\eta = 1 + \eta_1 + \eta_{2}(x/\theta_{e})^{\eta_3}$ is the Compton energy enhancement factor, and 
%-----------------------------------------------------------------
\begin{equation}
\begin{split}
    &x = \frac{h\nu}{m_{e}c^2}, 
    x_c = \frac{h\nu_c}{m_{e}c^2},\\
   &x_1 = 1 + 4\theta_e + 16 \theta_{e}^{2},
    x_2 = 1 - \rm{exp}(- \tau_{es}),\\
    &\eta_1 = \frac{x_2(x_1 -1)}{1 - x_1 x_2},
    \eta_2 = \frac{-eta_1}{3^{\eta_3}},
    \eta_3 = -1 - \frac{\rm{ln}x_2}{\rm{ln}x_1},
\end{split}
\end{equation}
%-----------------------------------------------------------
where $h$ is Planck’s constant and $\nu_c$ is the critical frequency, below which it is assumed that the emission is completely self-absorbed and above which the emission is assumed to be optically thin.

\section{Links to observations in XRBs}
\label{sec3}

X-ray binaries (XRBs) are the brightest compact sources in galaxies and have been used as unique tools to study not only the accretion onto a compact object, but also the General Relativity in the strong gravitational field regime (see reviews from \cite{Zhang2013, Belloni2016}). Moreover, Galactic black hole binaries (BHB) provide prototypes for the supermassive black holes in AGNs, with the advantages that stellar-mass BHs have much higher flux and much shorter characteristic timescales than AGNs.

\subsection{Spectral components and identifications}
\label{sec3:1}

% multi-temperature BB from a thin disk (i.e. radial temperature profile, optically thick case, some basic equations)
%- accretion states and origin of more complex spectral shapes
%- reflection and iron line diagnostics

The contributions from the optically thick and optically thin emission mechanisms can be easily identified in the observed spectra of X-ray binaries (BHBs) as soft and hard spectral components. During an outburst, the relative strengths of these components change frequently in concert with the changes in luminosity, thereby develops a set of empirical spectral classification states that can be used to broadly characterize the underlying physical state of accretion flow.

\subsubsection{Accretion disk}
\label{sec3:1:1}

The soft component is believed to originate from the radiation of the SSD, which is indeed confirmed by the observed $L_{\rm{disk}} \propto T_{d}^4$ relation between the disk luminosity and disk temperature in BHBs \cite{Davis2006, Dunn2011}. The theoretical spectrum of SSD in the optically thick limit is a sum of blackbody spectra of different temperatures (Section \ref{sec2:4:1}, Eq. \ref{eq.L_disk} -- \ref{eq.:T_disk}). 

In realistic situations, a widely applied model to interpret observations of BHBs is the so-called 'multicolor disk blackbody model' (known as \textit{diskbb} in \textsc{XSPEC}) \cite{Mitsuda1984}, assuming that the viscous torque vanishes at the ISCO, thereby allowing easy integration of the total flux. The disk temperature Eq. \ref{eq.L_disk} at the inner disk radius $r_{\rm{in}}$ is given at
%------------------------------------------------------
\begin{equation}
\label{eq.:T_in}
 T_{\rm{in}} = K \left [ \frac{3GM\dot{m}}{8\pi\sigma r^3_{\rm{in}}}(1-\sqrt{\frac{6R_{\rm{g}}}{r_{\rm{in}}} } ) \right ] ^{1/4},
\end{equation}
%--------------------------------------
while the peak of the energy spectrum can be found at $\sim f_c 2.36T_{\rm{in}}$, where $f_c$ is the color correction. To improve the accuracy of the spectral fittings, a number of modified blackbody models has been proposed to include different inner viscous torque boundary conditions, Doppler effect due to the rotation of the matter in the disk, Thomson scatterings in the upper layer and in the atmosphere of the disk, etc \cite{Ebisawa1991, Shimura1995, Ross1996, Davis2006}.

Therefore, in principle, a well-described spectral energy distribution for the emission produced by a SSD around a black hole of known mass, could put constraints on the overall radiative efficiency of the accretion process. More importantly, it puts constraints on the inner boundary condition of the accretion disc, and therefore on the black hole spin \cite{Zhang1997}. Particularly, the BH spin changes the location of ISCO and hence changes the effective inner edge of the disk, which ultimately changes the $T_{\rm{in}}$. The higher the spin is, the smaller the ISCO is, and the higher the inner disk temperature is. Note, the observed thermal spectrum also depends on the disk inclination, thus this parameter must be determined before BH spin can be extracted from the thermal spectrum (see \cite{Reynolds2021}). 

\subsubsection{Corona}
\label{sec3:1:2}

\begin{figure}
\includegraphics[width=11cm]{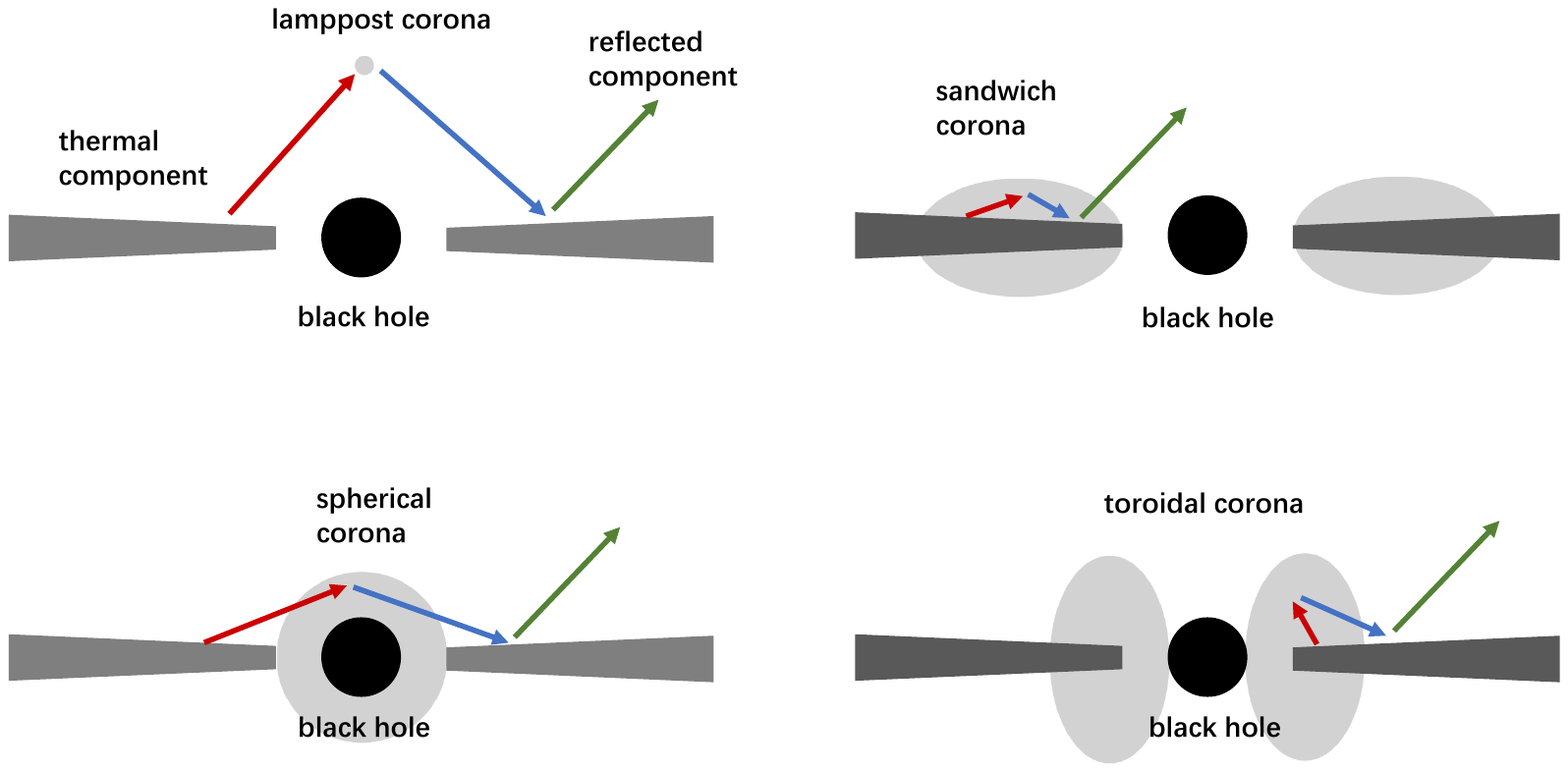} 
\caption{Examples of possible corona geometries: lamppost geometry (top left), sandwich geometry (top right), spherical geometry (bottom left), and
toroidal geometry (bottom right). Figure reproduced from Bambi (2017) \cite{Bambi2017}.}
\label{fig3} 
\end{figure}

The hard component is generally thought to originate in corona through Compton up-scatterings of the soft photons from accretion disk or the synchrotron emission of the electrons \cite{Galeev1979, Poutanen2018}. Despite decades of numerous studies, there still no consensus on the detailed geometry of the corona. For example, the 'lamppost' model assumes that the corona locates on the rotation axis of the BH, above a standard thin disk that extends to the ISCO, which may also correspond to the base of the jet \cite{Markoff2005, Markoff2010}; the 'sandwich' model proposes that the corona is the atmosphere above the inner part of an optically thick accretion disk \cite{Beloborodov1998a}; the ADAF configuration (spherical corona) assumes that corona is the inner hot flow that lies between the inner radius of accretion disk and the BH, meaning that the accretion disk has to to truncated at certain radius, etc. Fig. \ref{fig3} shows examples from some plausible configurations of the corona. Moreover, recent studies have suggested that the geometry of corona evolves with time \cite{Kara2019} and could be in-homogeneous \cite{Mahmoud2018a, Mahmoud2018b, Garcia2021, Yang2022}. 
%----------------------------------------------------
%----------------------------------------------------  

Nevertheless, it is clear that, the electrons in the corona can not be uniformly thermal due to the heating of the accretion disk, which consequently affects the shape of the observed hard spectrum (i.e, the value of the photon index $\Gamma$) \cite{Poutanen2013}. The photon index $\Gamma$ of the Comptonized spectrum depends on the parameters of the Comptonizing matter, primarily on the electron temperature, $T_e$, and the Thompson optical depth $\tau_e$ \cite{Sunyaev1980, Poutanen1996}, 
%----------------------------------------------------------------
\begin{equation}
    \Gamma = \sqrt{\frac{9}{4} + \frac{\pi^2 m_e c^2}{3\kappa_{\rm{B}} T_e (\tau_e + 2/3)^2}} -1/2.
\end{equation}
%---------------------------------------------------------------------
The Comptonized spectrum cuts off exponentially at higher energies when the electrons reach an approximate equipartition with the up-scattered photons. Sometimes, in the soft state of BHBs, the electrons in the corona also contains a high energy non-thermal tail beyond the thermal distribution \cite{Poutanen2013, Gierlinski2003, Zdziarski2017}. 

%A non-thermal tail can affect the reflection spectrum at high energies, where reflection is purely Compton scattering \cite{Bambi2021}. 

\subsubsection{Reflection}
\label{sec3:1:3}

Besides the thermal and Comptonized components, the other commonly observed component in the X-ray spectra of accreting BHs is the reflection emission. The typical features of X-ray reflection component are the relativistic broadened Fe-K$\alpha$ emission line at $\sim$ 6.4 keV and the Compton hump peaked around 20--30 keV. 

X-ray hard photons from the corona can be down-scattered by the accretion disk and further reprocessed to produce a reflection component. The Compton hump results from the combination of photo-electric absorption of low-energy photons and multiple electron down-scattering of the hard photons. Since the reflection mainly comes from the innermost regions of the accretion disk, the reflection spectra studies of XRBs can be a powerful tool to study the region close to BHs. 

The observed X-ray reflection spectrum is distorted by the Doppler effect and the gravitational redshift of the black hole potential. Both two effects become stronger as accretion disk approaching the BH. The X-ray reflection spectra has been widely used to measure the spin of the BH. This method assumes that the X-ray reflection spectrum is truncated by the ISCO, thus both the ISCO and spin can be measured via the strength of the Doppler and gravitational broadening. 

The application of the X-ray reflection spectroscopy (RS) method works on black hole X-ray binaries in their luminous hard states, with typical accretion rates of $\dot{m} \sim 0.01 -0.3 \dot{m_{\rm{Edd}}}$. However, the difficulties lie in properly estimating the extent of the 'red' wing, which is most directly related to the spin of the black hole, and in modeling the hard X-ray source photons and the disk ionization, both of which strongly affect the reflection spectrum \cite{Reynolds2021}.

\subsubsection{Spectral states}

\begin{figure}
\includegraphics[width=55mm, height=45mm]{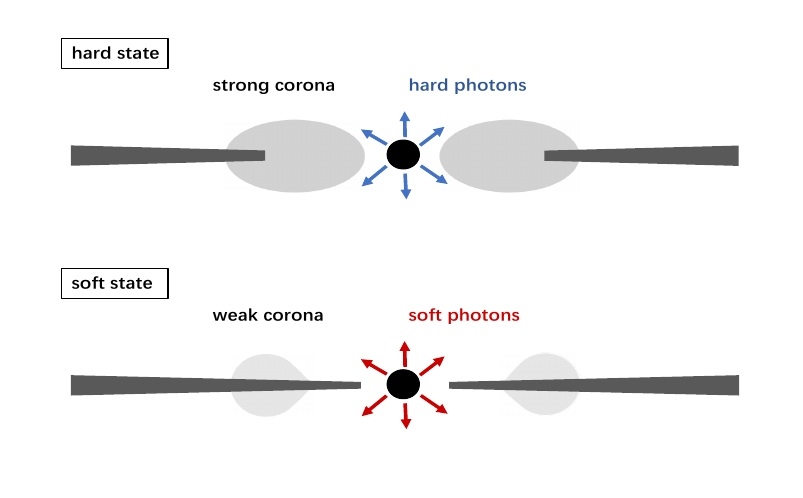}
\includegraphics[width=55mm, height=45mm]{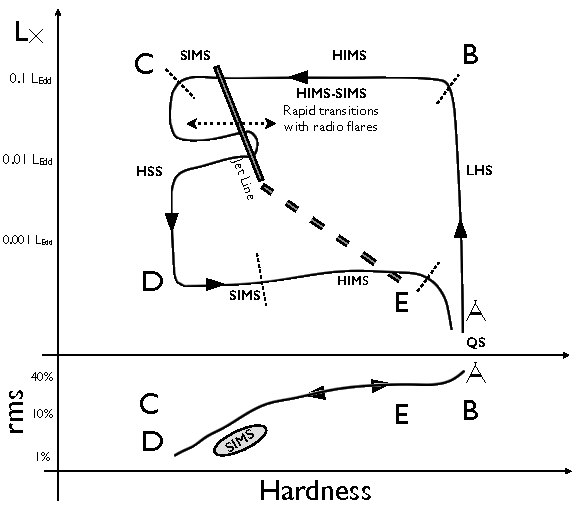} 
\caption{\textit{Left panel}: Description of the accretion flow structures in the historical hard and soft states. \textit{Right panel}: A schematic description of HID and HRD of a BHB outburst. The letters refer to main locations are described in the text. Figure adapted from Belloni (2016) \cite{Belloni2016}.}
\label{fig4} 
\end{figure}

Conversely, the spectrum of XRBs is a combination of the emissions from the disk, corona and the reflection. During an outburst of a BHB, the source exhibits different spectral states and transitions, in particular, the X-ray spectral state transitions between the historical high/soft state (HSS) and low/hard state (LHS, see \cite{Belloni2016} for a review). Typically, in the LHS, the spectrum of BHBs is dominated by a power-law component ($\Gamma \sim$ 1.5--2.1) in hard X-rays, which is thought to be produced by the Compton scattering of soft photons off the hot electrons in the corona. In the HSS, the spectrum is characterized by a multi-color blackbody component that dominates at about 1\ keV, which is believed to arise from a geometrically thin, optically thick accretion disk, covered by a weak/hot corona. The typical accretion geometry is shown in the left panel of Fig. \ref{fig4}.
%-----------------------------------------------------------------
%----------------------------------------------------------------------

Recently, the observations of BHBs have shown more complicated spectral features, suggesting a combination of various accretion flows \cite{Kara2019, You2021, Ruan2019}, accompanied by wind/outflows in some cases (see \cite{Yuan2015, Bu2016, Bu2018} and references therein). Nevertheless, a common accepted scenario of the accretion geometry in all these observations is the co-existence of hot and cold accretion flows, of which the systems have either an inner ADAF connecting to a truncated disk, or a 'lamppost" corona lying above a standard thin disk which extends inward to the ISCO. 

Most BHBs are transients, spending most of their lifetimes in the quiescent state (QS), and occasionally go into an outburst that could last from a few days to several months (except for one peculiar source, GRS 1915+105, has been active since the discovery). Despite that the outbursts of different systems have variant spectral-timing properties, these systems generally show similar behaviors in a Hardness-Intensity Diagram (HID) and a Hardness-Rms Diagram (HRD) \cite{Belloni2016}, with the hardness defined as the ratio of count rates between a soft band and a hard band. From the HID (Fig. \ref{fig4}, left panel), one can easily identify the two states as the two 'vertical' branches. The right branch corresponds to the LHS, which is observed at the start and at the end of an outburst only, with relatively low accretion rate. The left branch corresponds to the HSS, with relatively high accretion rate. In addition, there are two intermediate states in the central part of the diagram, recognized as the Hard Intermediate State (HIMS) and Soft Intermediate State (SIMS). The precised transition between these four states need additional information on the properties of fast variability and/or changes in the multi-wavelength relations \cite{Belloni2020}.

Despite various accretion solutions have been proposed to explain the accretion flow of BHBs, the 'truncated disk model' might be the most successful one in explaining the disk truncation and spectral states transition. This model pictures the hard state as a truncated SSD thin disk adjoined with an inner ADAF flow (or corona) that lie between the truncated disk and the ISCO. At this stage, the disk is truncated at very large radii, only a few photons from the disk illuminate the flow, the spectra show typical shape of a thermal Comptonization distribution. As the disk moves inwards, it extends further underneath the flow, so that there are more seed photons intercepted by the flow, leading to a softer spectra. This scenario gives us a spectrum that is a combination of both hard and soft components and a thin disk with a progressively smaller inner radius. These are the typical features of the spectra observed during the (usually) short-lived HIMS and SIMS. When the truncated disk reaches the ISCO, the hot flow collapses into a SSD, where the spectrum is characterized by a multicolor-blackbody component.

\begin{figure}
\includegraphics[width=11cm]{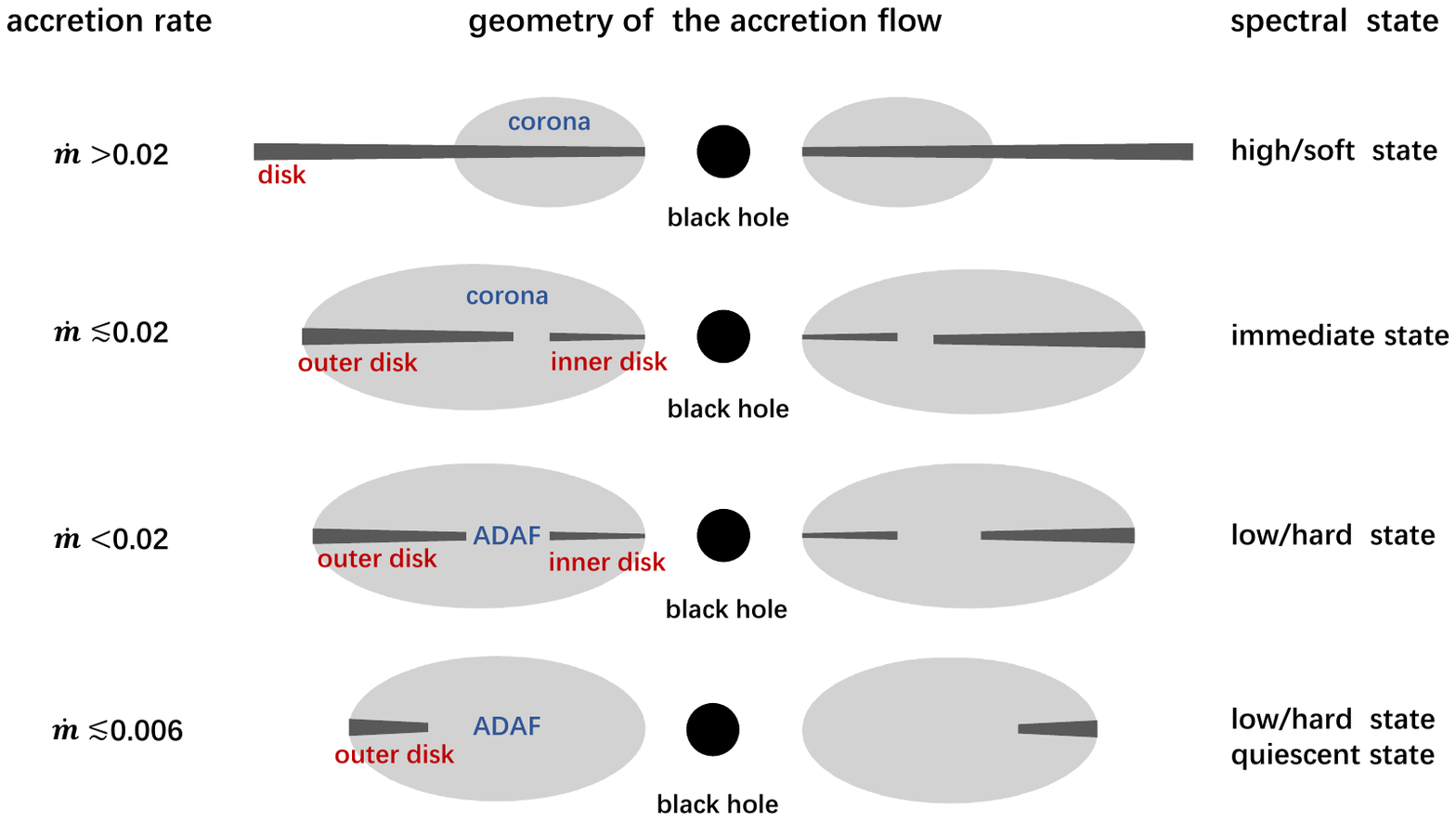} 
\caption{A schematic description of the accretion flow structures in different spectral states as consequences of disk-corona interaction, driven by mass accretion rate $\dot{m}$. Figure reproduced from Liu (2022) \cite{Liu2022}.}
\label{fig5} 
\end{figure}

However, the physical mechanism triggering the truncation of disk is still under discussion (see the review from \cite{Liu2022}). Among various possibilities, the disk and corona coupling model seems to be a promising scenario (Section \ref{sec2:4:4}). Specifically, the interaction between the disk and corona causes disk gas evaporating to the corona or coronal gas condensing to the disk, depending on the gas supply rate and how the gas feeds to the accretion. The evolution of accretion geometry during an outburst of BHBs is illustrated in Fig. \ref{fig5}.

%-----------------------------------------------------------------
%----------------------------------------------------------------------

\subsection{Timing perspectives on accretion}
\label{sec3:2}
%- general considerations (timescales and variability)
%- perturpation/propagation model for broadband noise
%- QPOs and their origin
%- LT precession
%- jets
%- time lags and coherence

%\subsubsection{Fast Variability}
%\label{sec3:2:1}

Besides the spectral properties, fast variability is also an important characteristic of accreting BHs and a key ingredient for understanding the accretion processes in these systems \cite{Motta2016, Ingram2019, Belloni2020}. The fast variability is believed to be produced by the in-homogeneity in the inner accretion flow. For a given radius, if the timescale of the in-homogeneities is longer than the orbital timescale, this could lead to a signal concentrated around the frequencies corresponding to that radius. Hence the in-homogeneity could be used as a 'test particle' to probe the geometry of accretion flow.

BHXBRs typically show fast X-ray variability on a wide range of timescales Fast variability is usually studied through the power density spectra (PDS) that is composed of broad-band noise and peak components called Quasi-Periodic Oscillations (QPOs). The strength of the fast variability is quantified as the fractional rms of the PDS. It is observed to have a very strong connection with the spectral state of BHBs, i.e., the accretion geometries of these systems. Generally, the PDS is dominated by the noise component during the LHS and HSS and is characterized with QPOs during the intermediate states. In some cases, the type of QPOs can even help to distinguish the intermediates states.

\subsubsection{Noise and propagation}
\label{sec3:2:1}

In the LHS, PDS of BHBs is dominated by strong band-limited noise with   fractional rms values up to 40--50 \%, with usually no QPOs. At this state, The energy spectrum is dominated by the corona component, which is high variable. The PDS can be described with a sum of Lorentzian functions that measure the characteristic frequencies of the main components seen on the PDS. Generally, there are four components that have been observed in a BHB: a low frequency flat-top breaks at certain value, a low-frequency bump (could evolve to a QPO), a broader bump at higher frequency and a highest frequency break. During an outburst, as source flux (and accretion rate) increases, the total rms decreases slightly, all characteristic frequencies increase and the energy spectrum softens. 

One of the most promising model proposed to incorporate these frequencies is the propagating fluctuation models \cite{Lyubarskii1997, Kotov2001}. In this model, the band-limited noise components are proposed to break down at local viscous frequency $f_{\rm visc} \propto 1/R^2$. Perturbation occurs at each radius of the accretion flow, but the fluctuation from the outer region will modulate the inner region because the inward motion of accretion flow. The outer and inner radii of this hot flow were suggested as the origins of the low and high frequency breaks (or broad Lorentzians)in the PDS. Generally, numerical models combining the propagating fluctuations process with a hot flow show widely agreement with the variability properties observed in XRBs. 

However, when considering only the variability in the hot flow, it turned out to be difficult to accurately reproduce all the observed timing properties with the model. Several improved propagating fluctuation models have been proposed by considering: a considerably variable disk that propagates variability into the hot flow \cite{Rapisarda2016}; an extra variability in the hot flow and different propagation speeds of the fluctuations \cite{Rapisarda2017}; an additional backward propagation from the hot flow \cite{Mushtukov2018}; or a spectrally in-homogeneous hot flow \cite{Mahmoud2018a, Mahmoud2018b, Kawamura2022}, in order to match the power spectral shape and width observed in BHs. However, these improved models recently have been challenged by the high energy variability observed at 30--200 keV \cite{Yang2022, Kawamura2023}, which encourages further investigation of the fundamental hypotheses of the propagating fluctuations model.  

On the contrary to LHS, in the HSS, variability is very weak and limited to a few \%, and the PDS typically has a pure power law shape. The energy spectrum is dominated by the thermal disk component, which is not much variable. However, this conclusion is restricted by the instrument capability, since most of the instruments are not sensitive to photons below 0.5\ keV. XMM-Newton studies on BHBs have shown that the disk variability is not only strong in the hard state, but also precede the hot flow variability \cite{Uttley2014}. They proposed that that the disc variations could even drive the harder X-ray variability at lower lower frequencies. At higher frequencies, where are closer to the hot flow, the hot flow variability dominates over the disk variability. 

%However, this need not be the case: mass accretion variations at higher Fourier frequencies in the disc could be filtered by the more extended disc emissivity profile to suppress the variability of the direct disc emission at these frequencies. 

\subsubsection{QPOs}

QPOs are not only a common characteristic in BHBs, but also observed from cataclysmic variables, through neutron star XRBs and enigmatic ultra-luminous X-ray sources, to AGNs (see the review from \cite{Belloni2020}. In addition, universal characteristic frequencies correlations found in these systems suggest that the QPOs probably have a similar origin, independent of the nature of the compact object \cite{Wijnands1999, Psaltis1999, Bu2015, Bu2017}. 

Low frequency (LF) QPOs are commonly observed in the intermediate states of the BHBs, when both considerable emissions from disk and corona are observed in the energy spectrum. The frequency of LFQPOs vary over a range of frequencies from $\sim$ 0.01 to 30 Hz and can be much higher in neutron star systems. LFQPOs are defined in three types (type-A, -B and -C) and their appearances generally relate to the spectral states. For instances, type-C QPOs in the HIMS, type-B QPOs in the SIMS and type-A QPOs in the HSS. These three types of QPO have been shown not to be the same signal, based on, in particular, the simultaneous detection of a type-B and a type-C QPO \cite{Belloni2016}. 

Several models have been proposed to explain the origin of type-C QPOs, which are generally based on two different mechanisms: instabilities in the accretion flow and geometric effects under general relativity (see the review by \cite{Ingram2019}. While it seems no doubt that type-C QPOs are caused by an azimuthally asymmetric geometric effect that favours a precession origin, there still no unified paradigm has been confirmed. Among various proposed models, the Lense-Thirring (LT) precession from either a inner hot flow or a small-scaled jet is widely accepted in explaining the type-C QPOs observed in BHBs \cite{Ingram2009, Motta2014, Ma2021}. Specifically, the inner hot flow model assumes a truncated disk geometry, of which the BH spin axis is assumed to be moderately misaligned with the rotational axis of the binary system. The QPO frequency is determined by the LT precession of the inner hot flow. The observed flux is modulated by Doppler boosting and solid angle effects, and the broad-band noise associated with QPOs arises from the variations in mass accretion rate from the outer regions of the accretion flow that propagate toward the BH, modulating the variations from the inner regions and, consequently, modulating also the radiation in an inclination-independent manner \cite{Ingram2013}. On the other hand, the jet-precession model assumes the LT precession comes from a small-scaled jet instead of the hot flow, which doesn't necessarily require a truncated disk \cite{Ma2021}. While the LT precession can in principle explain the type-C QPO properties upon the source inclination, recent results appear to challenge the model, at least in its current form \cite{Marcel2021, Nathan2022}, which further arises fundamental questions on whether the LT precession radius is the inner disc radius or whether the LT precession is the ``right" precession for type-C QPOs. Recently, several improved disc–corona coupling models have been proposed to explain the radiative properties of the variability \cite{Karpouzas2020, Bellavita2022, Zdziarski2021, Kawamura2022}. Among these models, the \textsc{vkompth} model, assuming the QPO arises from the coupled oscillation between the corona and the disc, has been tested to be very promising in explaining the QPOs observed in XRBs, including the kilohertz (kHz) QPOs in neutron star system and type-A, -B, -C QPOs in BHBs \cite{Zhang2023, Garcia2021}.  

Type-B QPOs appear only in the SIMS and is found in narrow range of frequencies, i.e., around 6 Hz or 1--3 Hz. Their PDS also show very weak red noise, which distinguish them directly from type-C QPOs. It is widely accepted that type-B QPOs are associated with the relativist jet \cite{Fender2009}, however, not clear how. Some work has suggested that type-B could also originate from the LT precession of the jet \cite{Liuhx2022}.

Type-A QPOs are the least studied type of QPO because of their small numbers ($\sim$10). They are observed in the HSS, right after the hard to soft transition when the overall variability is already very low. One plausible model for this signal is based on the ``accretion ejection instability" (AEI), according to which this instability could form low azimuthal wave numbers driven by magnetic stresses, standing spiral patterns which would be the origin of type-A QPOs (see \cite{Motta2016} for a review). 

%Most recently results from MAXI J1348-630 suggested that type-A QPOs could also be caused by the coupled oscillations between the corona and the disc \cite{Zhang2023}.

Compared to LFQPOs, the high frequency (HF) QPOs in BHBs are much less studied due to much fewer detection, thus makes the tests of model difficult. They appear only when the accretion rate is high and can sometimes be seen in pairs. Because of their higher frequency, it is assumed to come from the innermost region of the disk, closer than where the LFQPOs are produced. They are proposed to be produced by the nonlinear resonance between orbital and radial epicyclic motion \cite{Belloni2016}. This resonance models have successfully explained HFQPOs with frequency ratio consistent with 2:3 or 1:2. 

\section{Conclusions}
\label{sec4}

%Since the early 1970s, our understandings on black hole accretion processes have increased significantly, from both theoretical modelling and observational applications. In current review, I intend to give a general picture of black hole accretion theory and its connections to observed phenomena such as spectral states and fast variability. 

There is no doubt that $\alpha$-prescription will continue to be the pillar of black hole accretion theory, not only because it allows one to build models that couple the dynamics and thermodynamics of the accretion flow phenomenologically just by a $\alpha$, but also it enables decades of theoretical work achieved valuable insights within this paradigm without addressing much fundamental physics. Decades of efforts has focused on the dynamics of the MRI turbulence, but very few on the thermodynamics that connects directly to the observations. For instances, it is still not clear what is the true nature of the thin disk transition radius that links to the observational characteristics of XRBs, i.e., the hard to soft transitions, the high frequency break noise, or the quasi-periodic oscillations. How do viscous instabilities affect the variability. What makes the accretion power distribute into various forms (jet, ejection, outflow). And most importantly, what is the nature of corona?

From the observational perspective, for the last decade, thanks to the availability of several new missions (Insight-HXMT, NICER, IXPE), our understandings on the BH accretion has enhanced significantly. These instruments provide enormous amount of data to test the growing number of theoretical models attempting to explain the spectra, timing and polarization of accreting compact objects. We are allowed to study the fast variability in much higher X-ray band, which directly puts new constraints on the structures of the disk/corona of XRBs. More interestingly, spectro-polarimetric studies seems very insightful in constraining the coronal geometry. Nevertheless, there are still many open questions (the origin and propagation of disk variability, the origin of HFQPOs, degeneracy in the spectral modelling, the location and structure of corona, etc.), waiting for new instruments capable of collecting more photons and reaching a much higher sensitivity for fast variations.

\begin{acknowledgement}

Q.C.B. acknowledges the support from the National Program on Key Research and Development Project (Grant No.2021YFA0718500) and the National Natural Science Foundation of China (Grant No. U1838201, U1838202, 11733009, 11673023, U1838111, U1838108 and U1938102).

\end{acknowledgement}

\clearpage
%\input{references}
%%%%%%%%%%%%%%%%%%%%%%%% referenc.tex %%%%%%%%%%%%%%%%%%%%%%%%%%%%%%
% sample references
% %
% Use this file as a template for your own input.
%
%%%%%%%%%%%%%%%%%%%%%%%% Springer-Verlag %%%%%%%%%%%%%%%%%%%%%%%%%%
%
% BibTeX users please use

%\bibliographystyle{acm}
%bibliography{}
%
%\biblstarthook{References may be \textit{cited} in the text either by number (preferred) or by author/year.\footnote{Make sure that all references from the list are cited in the text. Those not cited should be moved to a separate \textit{Further Reading} section or chapter.} If the citation in the text is numbered, the reference list should be arranged in ascending order. If the citation in the text is author/year, the reference list should be \textit{sorted} alphabetically and if there are several works by the same author, the following order should be used:
%

\end{document}